\documentclass[12pt]{article}

\usepackage{amsmath}
\usepackage[dvips]{graphicx}

\newcommand{\bea}{\begin{eqnarray}}  
\newcommand{\eea}{\end{eqnarray}}  

\baselineskip
\newcommand {\dperp} {d_{T}}
\newcommand {\hc} {{\not \!h}c}

\begin{document}

\begin{flushright}
hep-ph/0107224\\
LTH 512\\
MC-TH-01-06\\
\end{flushright}

\begin{center}
\vspace*{2cm}

{\Large {\bf Colour dipoles and virtual Compton scattering}} \\

\vspace*{1cm}

M.~McDermott$^1$, 
R.~Sandapen$^2$ and G.~Shaw$^2$

\vspace*{0.5cm}
$^1$Division of Theoretical Physics,\\
Department of Math. Sciences,\\
University of Liverpool.\\
Liverpool, L69 3BX. England. \\

\vspace*{0.2cm}
$^2$Department of Physics and Astronomy,\\
University of Manchester,\\
Manchester. M13 9PL. England.

\end{center}
\vspace*{2cm}

\begin{abstract}

\noindent

An analysis of  Deeply Virtual Compton Scattering (DVCS) is made within the colour dipole model. 
We compare and contrast two models for the dipole cross-section which have been successful in 
describing structure function data. Both models agree with the available cross section 
data on DVCS from HERA. We give predictions for various azimuthal angle asymmetries in HERA 
kinematics and for the DVCS cross section in the THERA region. 

\end{abstract}


\newpage

\section{Introduction}
\noindent
In this paper we explore the predictions of the colour dipole model for
high energy deeply virtual Compton scattering (DVCS):
\begin{equation}
\gamma^*(q) + p(P) \rightarrow \gamma (q') + p(P')  ,
\label{DVCS}
\end{equation}
where the first photon has spacelike virtuality $q^2 = -Q^2 > 0$, 
but the second photon is real ($q^{\prime 2} = 0$), and hence transversely polarized. DVCS is a particular example of a diffractive process
\begin{equation}
\gamma^*(q) + p(P) \rightarrow X + p(P')  \, ,
\label{gendiff}
\end{equation}
\noindent in which the diffractively-produced system, X, is separated by a 
rapidity gap from the elastically-scattered proton (at least for high photon-proton 
centre-of-mass energies, $W$, i.e.  $W^2 = (q + P)^2 \gg Q^2, M^2_{X}$). The
first HERA data on this process is now available \cite{zeusdata,h1data}.

The colour dipole model of diffraction \cite{dipole1} provides a simple 
unified picture of such diffractive processes and enables ``hard'' 
and ``soft'' physics to be incorporated in a single dynamical framework. 
At high energies, in the proton's rest frame, the virtual photon fluctuates 
into a hadronic system (the simplest of which is a $q {\bar q}$ {\it dipole}) a 
long distance upstream of the target proton. The formation time of this hadronic 
system, and of the subsequent formation of the hadronic final state, is much longer 
than the interaction time with the target. It is this observation that leads to the main (plausible) 
assumption of the colour dipole model, i.e. that the interaction of a given 
fluctuation with the target is independent of how it is formed, and is 
therefore universal. It leads to the following generic factorization of the amplitudes 
of high energy diffractive processes:
\begin{equation}
{\cal A} (\gamma + p \rightarrow X  p) = \int ~\psi^{\mbox{in}}_{\gamma} ~{\hat \sigma} ~\psi_{X}^{\mbox{out}}  
\end{equation}
where ${\hat \sigma}$ is the interaction cross section of a given configuration with the target and the integral 
runs over the phase space describing the incoming and outgoing hadronic systems. For the case of dipole scattering, one must 
integrate over dipole configurations (longitudinal momentum fractions and transverse sizes). 
We know of no formal proof of this type of high-energy factorization, whether applied to dipoles or more complicated configurations. 
Nevertheless, within this common framework there are many different formulations
for the interaction cross section ${\hat \sigma}$ \cite{levin1}-\cite{Heidelberg}, which have been applied with 
varying degrees of success\footnote{For a recent overview, see \cite{amirim}.}. Here we consider two particular dipole models \cite{FKS1,MFGS1} which have both been successful in describing structure function data, but which
at first sight differ quite drastically in their structure and implications; 
and compare their predictions for DVCS.

An additional assumption of most dipole models of diffraction is that the scattering with the target is 
diagonal with respect to the appropriate variables (i.e. transverse sizes, momentum fractions and polarizations are unchanged by the interaction). For the case of DVCS this implies that the incoming photon must be transversely polarized in order to respect s-channel helicity conservation.

DVCS is a good probe of the transition between soft and hard regimes in 
the dipole model for two reasons. Firstly, the transverse photon 
wave function can select large dipoles, even for large $Q^{2}$, and certainly 
for the $Q^2$ range $2 < Q^2 < 20$ GeV$^2$ for which data is 
now available \cite{h1data}. Secondly, because the final photon is real,
DVCS is more sensitive to large dipoles than DIS at the same $Q^2$, as we
shall illustrate quantitatively in Section 3.
In addition, for $Q^2 \rightarrow 0$, the process reduces to real Compton
scattering and the cross-section can be reliably inferred from real 
photo-absorption data, where soft physics dominates.

We stress the potential importance of well-founded dipole descriptions
in providing reliable starting points for exploiting DGLAP evolution
properties at ``large'' $Q^2$. From the theoretical point of view, DVCS is the best understood of 
all exclusive diffractive processes, essentially because the X system is just a real photon. 
Indeed, a perturbative QCD factorisation theorem has been explicitly
proven as $Q^2 \rightarrow \infty$ \cite{jcaf} which enables the 
QCD amplitude to be described by a convolution in momentum fraction of generalised (or skewed) parton distributions \cite{G2} (GPDs) 
with hard coefficient functions. GPDs correspond to Fourier transforms of  operator products evaluated 
between proton states of unequal momenta (cf. eq.(\ref{gendiff})). They are therefore generalizations of the familiar parton distributions of deep inelastic scattering, and like them satisfy perturbative evolution equations \cite{dglap, erbl, G3, bmns} which enable them to be evaluated at all $Q^2$ in terms of an assumed input at some appropriate $Q^2 = Q_0^2$. In practice, to compare with experimental results at finite $Q^2$ one must establish a regime in $Q^2$ in which the higher twist corrections (see e.g. \cite{bkms,ht}) to this leading twist result are numerically unimportant. 
This is a very difficult task in general but early indications are that the minimum $Q^2$ values defining this regime are considerably higher than in 
inclusive cross sections, for which values as low as $Q_0^2 = 1$~GeV$^2$, or even lower, have been used. Since the contributions from different transverse sizes are manifest in the dipole model one may realistically hope to gain insight into this question by investigating DVCS in the dipole framework. 
As such a dipole analysis of DVCS provides a complementary description to 
the formal QCD analysis, applicable at  ``large'' $Q^2$. Any insight 
gained regarding the mixture of soft and hard physics within the dipole model 
framework, can also be employed in those processes for which factorization theorems  
have not been proven.

Frankfurt, Freund and Strikman \cite{FFS} have given  a leading order QCD 
 analysis of DVCS. The resulting predictions for the DVCS amplitude at $t=0$ 
are in agreement with the recent 
 H1 measurements \cite{h1data} of the total DVCS cross-section,
 assuming an exponential t-dependence with a reasonable 
value of the slope parameter. The GPDs are evolved from an input 
value $Q_0^2 = 2.6$ GeV$^2$, where
the input GPDs are obtained by estimating their ratio to ``ordinary'' parton 
distribution functions (PDFs) using a simple aligned jet model\footnote{For
 further discussion of this approximation, see \cite{DGS1} and the original
paper \cite{FFS}.}.  
While this provides a reasonable first estimate, it is clearly subject to 
uncertainties which will become important when more accurate data are 
available.

Recently NLO QCD analyses of DVCS have been completed \cite{bmns, afmm1, afmm2, afmm3} which use as input GPDs Radyushkin's model \cite{rad} based on Double Distributions proportional to PDFs (which automatically impose the correct symmetry properties in the so-called ``ERBL region''). 
The colour dipole model offers a means of estimating these   
distributions at the input scale in the DGLAP region, in a complementary and well-founded framework, which can accurately describe both virtual 
Compton scattering and other closely related data over a wide range of $Q^2$. 
This is possible because at leading-log accuracy in $Q^2$, the amplitude is 
approximately equal to the GPD, at a particular point.
In this paper we compute predictions for the cross section of eq.(\ref{DVCS}), and for various azimuthal
angle asymmetries for the associated lepton process \cite{bmns2} which are
sensitive to both the real and imaginary parts of the DVCS amplitude.

The structure of the paper is as follows: in section \ref{sec:cdm} we summarize
and compare our two dipole models in the context of deep inelastic scattering;
we then discuss their application to virtual Compton scattering process in section \ref{sec:dvcsfr}; 
compute various  observables in section \ref{bh}  
and summarize our results and conclusions in section \ref{sec:res}.

\section{The colour dipole model}

\label{sec:cdm}

\begin{figure}[htbp]
  \begin{center}
   \includegraphics[width=10cm,height=3cm]{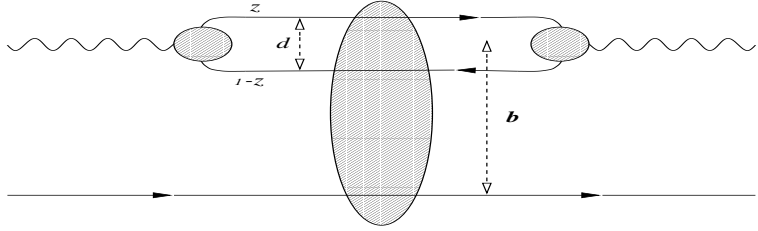}    
    \caption{The colour dipole model for the elastic process 
$\gamma^* p \to \gamma^* p$(DIS), and virtual Compton scattering
$\gamma^* p \to \gamma  p$. }
    \label{fig:fluct}
  \end{center}
\end{figure}

Singly dissociative diffractive $\gamma p$ processes (cf. eq.(\ref{gendiff}))
are conveniently described in the rest frame of the hadron, in which the 
incoming photon dissociates  into a $q \bar{q}$ pair
a long distance, typically of order of the ``coherence length'' $1/Mx$, 
from the target proton.  Assuming that the resulting partonic/hadronic 
state evolves slowly compared to the timescale of interaction with the proton or nuclear target,
it can be regarded as frozen during the interaction. In the  colour dipole
model, the dominant states are assumed to 
be $q \bar{q}$ states of given transverse size, $\dperp$.  
Specifically 
\begin{equation}
  \label{eq:photon_wf}
  |\gamma_{r} \rangle = \int \mbox{d}z ~\mbox{d}^{2} \dperp
~\psi^{\gamma}_{r} (z,\dperp,Q^2) \, |z,\dperp, \rangle + \ldots \; ,
\end{equation}
where $z$ is the fraction of light cone energy carried by the quark and
$\psi^{\gamma}_{r} (z,\dperp, Q^2)$ is the \emph{light cone wave function} of the
photon of polarization $r = T,L$. Assuming that these dominant states
are scattering eigenstates (i.e. that $z, \dperp$ and the quark helicities, which are
left implicit in the above equation, remain unchanged in diffractive 
scattering) 
the elastic scattering amplitude for $\gamma^* p \to \gamma^* p$ is 
specified by Fig. \ref{fig:fluct}. This leads via the optical theorem to  
\begin{eqnarray}
 \sigma^{\gamma^{*} p}_{T,L} &=& \int \mbox{d}z \mbox{d}^{2} \dperp \ |\psi^{\gamma}_{T,L}(z,\dperp,Q^2)|^{2} 
{\hat \sigma} (s^*,\dperp,z) \; ,
\label{absorptive} 
\end{eqnarray}
for the $\gamma^* p$ total cross-section in deep inelastic scattering,
where  ${\hat \sigma} (s^*,\dperp,z)$ is the total cross-section for scattering dipoles of specified  $(z,\dperp)$ which do 
not change in the interaction (the second line then follows from orthogonality and the variable $s^*$ will be specified shortly).

The dipole cross-section is usually assumed to  be flavour independent and``geometric'', i.e. 
independent of $z$.  Beyond this the models fall into two main classes.

In the first, the dipole cross-section is assumed to depend solely on the 
properties of the 
dipole-proton system itself, implying the choice $s^* = W^2$.  
Other singly diffractive photo-processes involve exactly the same dipole 
cross-section, but different wavefunction factors depending on the final 
state, as we shall
see below for virtual Compton scattering. 

The second type is more closely connected with hard perturbative QCD
predictions for the interaction cross section, which for 
small $x$ and high $Q^2$ involve two gluons being exchanged. For hard 
scattering, small
dipoles are connected via two parton lines to the proton. The interaction cross section then depends on the momentum fractions of the proton carried by the parton lines, i.e. 
$s^* = x_{Bj} = Q^2/W^2$, or $s^* = x' \approx x_{Bj}$.
In this case the dipole cross-section must be slightly modified when applying it to 
different processes, since GPDs of the appropriate kinematics must be used, as discussed below.  

In the rest of this section, we shall briefly summarize the properties of one model of each type, 
after first considering the other main ingredient in eq.(\ref{absorptive}), 
i.e. the photon wavefunction.

\subsection{The photon wavefunction}

Because the proton structure function, $F_2$, is predominantly transverse, 
both small and large dipoles contribute significantly to $F_2$ over a wide 
range of $Q^2$, where
``large'' means transverse sizes of order $\dperp \approx 1$ fm. With this 
caveat,
it is none the less useful to consider two qualitatively different regimes.

For small dipoles,  it is reasonable to
assume  ``QED wavefunctions,'' $\psi_{T,L}^{\gamma} = \psi_{T,L}^0$
calculated\footnote{For an explicit derivation, see Appendix A of 
\cite{DGKP}.}  from the usual QED vertex $-i e \gamma^{\mu}$ . 
Explicitly
\begin{eqnarray}
  \label{eq:psi^2}
  |\psi_{L}^0(z,\dperp, Q^2)|^{2} & =  & \frac{6}{\pi^{2}}\alpha_{e.m.} \sum_{q=1}^{n_{f}}e_{q}^
{2} Q^{2} z^{2} (1-z)^{2} K_{0}^{2}(\epsilon \dperp) \\
  |\psi_{T}^0(z,\dperp, Q^2)|^{2} & = & \frac{3}{2 \pi^{2}}\alpha_{e.m.} \sum_{q=1}^{n_{f}}e_{q}^
{2} \left\{[z^{2} + (1-z)^{2}] \epsilon^{2} K_{1}^{2}(\epsilon \dperp) + m_{q}^{2} 
K_{0}^{2}(\epsilon \dperp) \right\}
\end{eqnarray}
where
\[
 \epsilon^{2} = z(1-z)Q^{2} + m_{q}^{2}\; ,
\]  
 $K_{0}$ and $K_{1}$ are modified Bessel functions and the sum is over quark
flavours. Furthermore, for 
the large $Q^2$ values where small dipoles dominate, these  wavefunctions 
become
insensitive to the quark mass. In this regime,
the wavefunctions are essentially known.

For small $Q^2~\frac{<}{~}~4 m_q^2$, the QED wavefunctions become 
sensitive to the squared quark mass $m_q^2$. At the same time, large dipoles, 
for which one would expect significant confinement corrections, become very 
important. The wavefunction is clearly model dependent in this region.

\begin{figure}[htbp]
  \begin{center}
   \includegraphics[width=10cm,height=6cm]{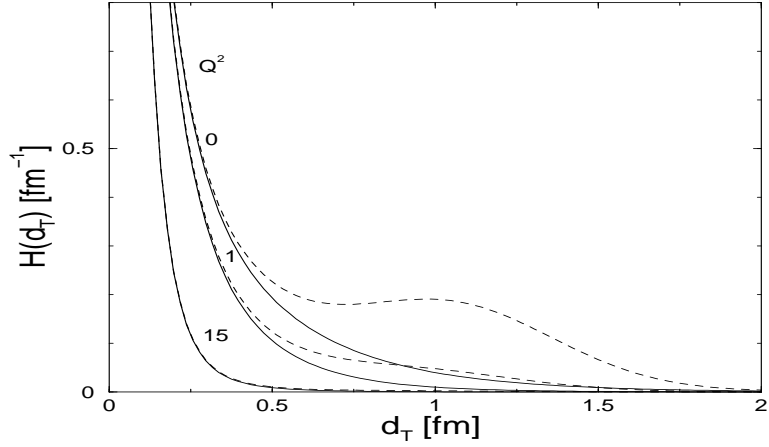}    
    \caption{The weight function $ H(\dperp)$ for different $Q^{2}$ 
corresponding to the photon wavefunction obtained by FKS \cite{FKS1} 
with $m_q^2 = 0.08$~GeV$^2$. The peak at low $Q^{2}$ represents the modification to 
the perturbative photon wave function at large $\dperp$.}
    \label{fig:fig2}
  \end{center}
\end{figure}
\begin{figure}[hp]
  \begin{center}
   \includegraphics[width=10cm,height=6cm]{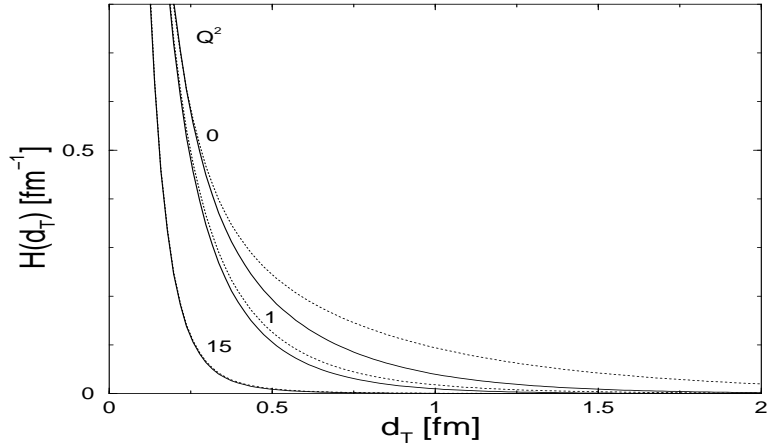}
    \caption{The weight function $ H(\dperp)$ for various $Q^2$  corresponding
to photon wavefunctions of perturbative form for $m_q^2 = 0.08$~GeV$^2$ (solid lines) and 
$m_q^2 = 0.02$~GeV$^2$ (dotted lines).}
    \label{fig:wfmdep}
  \end{center}
\end{figure}

Both the models discussed in this paper assume that the dipole cross-section 
becomes 
``hadron-like'' for $\dperp \approx 1$ fm, with an energy dependence 
characteristic of 
the ``soft Pomeron''. In choosing the
wavefunction in this region, FKS \cite{FKS1} were motivated by the work 
of Frankfurt,
Guzey and Strikman \cite{FGS}. These authors analyzed the distribution of
 scattering  eigenstates in a
non-diagonal generalized vector dominance model \cite{GVD1} which 
provides a good 
description of the soft Pomeron 
contribution to the nucleon structure function $F_2$ on both protons and
nuclei \cite{GVD2}. They found a distribution of states which was 
qualitatively similar 
to that obtained in a colour dipole model with a perturbative wavefunction,
but with an enhanced contribution from dipole cross-sections of hadronic size. 
In the light of this, FKS chose $m_q^2 = 0.08$ GeV$^2$ corresponding roughly
to a constituent mass for the light quark case; and modified the QED 
wavefunction by multiplying by
 an adjustable Gaussian enhancement factor:
\begin{equation}
  |\psi_{T,L}(z,\dperp, Q^2)|^{2} = |\psi_{T,L}^0(z,\dperp, Q^2)|^{2} \, f(\dperp)
\label{peak1}
\end{equation}
where
\begin{equation}
  f(\dperp) = \frac{1 + B \exp(- c^{2} (\dperp - R)^{2})}{1 + B \exp(- c^{2} R^{2})}.
\label{peak2}
\end{equation}
This form enables the width and height of the enhancement to be controlled 
independently while keeping a factor of close to unity at small $\dperp$. 
The effect of this is 
conveniently summarized by integrating
out the angular and $z$  dependence in eq.(\ref{absorptive}) to  give
\begin{eqnarray}
\label{eq:z_int}
\sigma^{\gamma^{*}p}_{tot} & = & \int \mbox{d}z \ \mbox{d}^{2} \dperp \
( |\psi_{T}(z,\dperp)|^{2} + |\psi_{L}(z,\dperp)|^{2} ) {\hat \sigma} (s,\dperp) \nonumber \\
   & = &  \frac{12}{\pi}\alpha_{e.m.} \int  
\mbox{d} \dperp H(\dperp) {\hat \sigma} (s,\dperp) \; . 
\end{eqnarray}
The resulting behaviour of $H(\dperp)$ for the final parameter 
values (see below) is shown in Fig. \ref{fig:fig2}: 
as can be seen, the enhancement is important for very low $Q^2$, but 
decreases rapidly as $Q^2$ increases.
Other authors do not in general include an explicit enhancement factor, but 
achieve a similar effect, at least for $Q^2 > 1$ GeV$^2$, by varying the 
quark mass. Choosing a smaller quark
mass increases the wavefunction at all large $\dperp$, as illustrated in
Fig. \ref{fig:wfmdep}. Golec-Biernat and Wusthoff \cite{GBW1}, for example, 
used  $m_q^2 = 0.02$ GeV$^2$, comparable with the pion mass squared. 

For our second model, the MFGS model \cite{MFGS1}, this question is less 
important, since results are only presented for $Q^2 > 1$~GeV$^2$; and 
results are presented for $m_q^2 = 0.08$~GeV$^2$ without a confinement factor of eq.(\ref{peak2}). In both models, a 
charmed quark contribution has also been included, which only differs from
the up quark contribution by the quark mass $m_c^2 = 1.4$~GeV$^2$.

Given this uncertainty in the wavefunction, it is clear from eq.(\ref{absorptive})
that the dipole cross-section at large $\dperp \approx 1$ fm cannot be 
inferred,
even in principle, from structure function and real photo-absorption data
alone. Other information must also be used. 

\subsection{The FKS model}

\begin{figure}[htbp]
  \begin{center}
   \includegraphics[width=10cm,height=6cm]{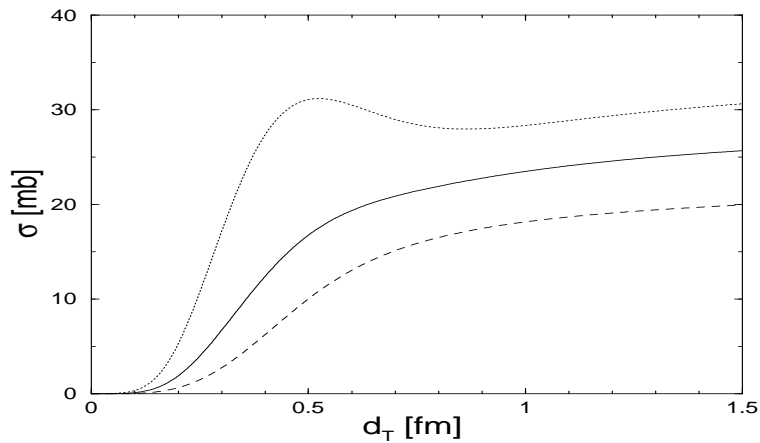}
    \caption{The FKS dipole cross section at $W = 10, 75, 300$~GeV (dashed, solid and dotted lines, respectively).} 
    \label{fig:fksdipole}
  \end{center}
\end{figure}

This model \cite{FKS1} belongs to the class in which  the 
dipole cross-section is assumed to depend solely on the properties of the 
dipole-proton system itself, implying the choice $s^* = W^2$  
independent of the virtuality of the incoming (or outgoing) photon. 
The idea was then to extract information on the dipole cross-section
by assuming a reasonable but flexible parametric form to fit  
structure function and real photoabsorption data in the 
diffractive region $ x \le 0.01$ for $0 \le Q^2 < 60$ GeV$^2$.
This was  implemented by assuming a sum of  two terms 
\begin{equation}
\label{gammatot}
{\hat \sigma} (W^2, \dperp)  =  {\hat \sigma}_{{\mbox{{\small soft}}}} (W^2, \dperp) +
{\hat \sigma}_{\mbox{{\small hard}}} (W^2, \dperp) \, ,  
\end{equation}
\noindent each with a Regge type energy dependence on the dimensionless energy variable $\dperp^2 W^2$:  
\begin{equation}
{\hat \sigma}_{\mbox{{\small soft}}} (W^2,\dperp)= a_{0}^{S} (1-\frac{1}{1+a_{4}^{S} \dperp^{4}}) (\dperp^{2} W^2)^{\lambda_{S}}
\label{sigmasoft}
\end{equation}
\begin{equation}
{\hat \sigma}_{\mbox{{\small hard}}} (W^2,\dperp)=(a_{2}^{H} \dperp^{2}+a_{6}^{H} \dperp^{6}) ~\exp (-\nu_{H} \dperp) (\dperp^{2} W^2)^{\lambda_{H}}
\label{sigmahard}
\end{equation}
These functions were chosen\footnote{In \cite{FKS1} a more complicated parametric form was used, but
this simpler parametric form gives very similar results.}
so that for small dipoles the hard term dominates yielding a behaviour 
$$
{\hat \sigma}  \rightarrow a^{H}_{2} \dperp^2 ~(\dperp^2 W^2)^{\lambda_H} \hspace{1cm} \dperp \rightarrow 0
$$
in accordance with colour transparency ideas. For large dipoles the soft term dominates with a hadron-like behaviour 
$$
{\hat \sigma}  \approx a_0^S ~(\dperp^2 W^2)^{\lambda_S} \hspace{1cm} \dperp \approx 1\; 
{\rm fm}   .
$$ 
The values $\lambda_S \approx 0.06$, $\lambda_H \approx 0.44$
resulting from the fit are characteristic of the soft and hard Pomeron
respectively, but the fits could be obtained for a range of values for the
parameter  $a_0^S$ because of the uncertainty in the photon
wavefunction at large $\dperp$ discussed above. This ambiguity can be resolved
by using the same dipole cross-section to calculate the structure function, 
$F_2^{D(3)}(x,Q^2, M_X^2)$, for diffractive deep inelastic scattering (DDIS)
\begin{equation}
\gamma^* + p \rightarrow X + p
\label{DDIS}
\end{equation}
and a subsequent paper \cite{FKS2} showed that good agreement was found for
$a^{S}_{0} \approx 30$~GeV$^{-2}$. The parameter values for this fit are
given in Table 1 and the resulting behaviour of the dipole cross-section as a 
function of $\dperp$ is shown in Fig. \ref{fig:fksdipole} for three
energies, including $W=75$~GeV corresponding
to the mean energy of the virtual Compton scattering data \cite{h1data} to 
be discussed below.

\begin{table}[htbp]
  \begin{center}
{\bf Table 1}
\[
    \begin{array}{c|c|c|c} 
\hline
 \lambda_{S}  & 0.06 \pm 0.01 & \lambda_{H} & 0.44 \pm 0.01 \\
            &              &              &    \\ 
 a_{0}^{S}    & 30.0 \mbox{ (fixed)} & a_{2}^{H}  & 0.072 \pm 0.010  \\
             &              &              &    \\
a_{4}^{S}    & 0.027 \pm 0.007   &  a_{6}^{H}   & 1.89  \pm 0.03 \\
             &              &              &    \\
             &              &  \nu_{H}  &  3.27 \pm  0.01 \\
             &              &              &    \\
B            & 7.05 \pm 0.08  &  c^{2}  & 0.20 \mbox{ (fixed)} \\
            &              &              &    \\ 
R            & 6.84 \pm 0.02  &          &  \\
 m_{q}^2 & 0.08  \mbox{ (fixed)} &  m_{c}^2  
                  & 1.4  \mbox{ (fixed)} \\
\hline  
    \end{array}
\]

    \caption{Parameters for the FKS model specified by eqs.(\ref{peak1}, \ref{gammatot}), 
in appropriate GeV based units throughout.}
    \label{tab:41.872}
  \end{center}
\end{table}

One feature of the FKS model is that, in its present form, it does not 
include ``gluon saturation'' or ``unitarity corrections'' which are expected to eventually damp the rapid rise with energy of the dipole cross-section for small dipoles. Its success implies
that such effects are not necessarily required in the HERA region. However, by examining
the predicted behaviour of the dipole cross-section, the authors have 
argued  \cite{FKS3} that saturation effects will begin to play a role
at the top of the HERA range and will rapidly become important above it.
This is illustrated in Fig. \ref{fig:fksdipole}, which  shows that the 
rapidly rising dipole cross-section at
small $\dperp$, where the hard term dominates, reaches hadronic sizes at 
the top of
the HERA range $ W \approx 300$ GeV.  
 
\begin{figure}[htbp]
  \begin{center}
   \includegraphics[width=10cm,height=6cm]{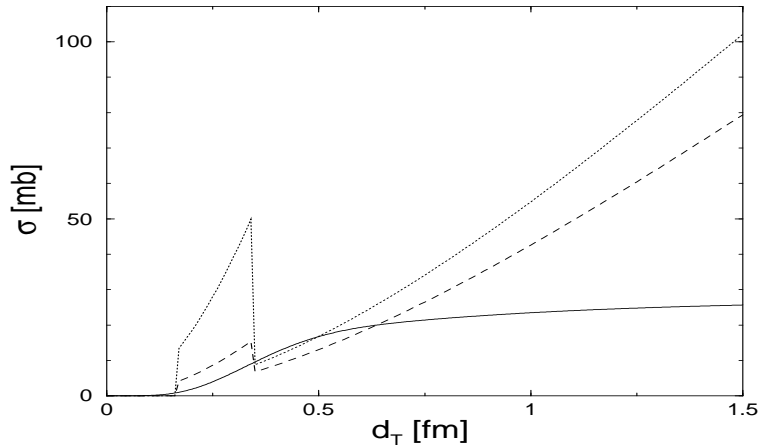}
    \caption{The Donnachie-Dosch dipole cross section \cite{Heidelberg} at 
$W =  75$~GeV (dashed line) and $W = 300$~GeV (dotted line). The FKS dipole
cross-section at $W =  75$~GeV (solid line) is included for comparison.} 
    \label{fig:dd_dipole}
  \end{center}
\end{figure}

Finally, for completeness, in Fig. \ref{fig:dd_dipole} we compare the 
FKS dipole cross-section
with that used in a recent analysis of DVCS by Donnachie and Dosch 
\cite{Heidelberg}. 
This model associates dipoles whose sizes are less than
(or greater than) a certain critical size with a fixed power energy
dependence corresponding to the hard (or soft) Pomeron respectively\footnote{
In contrast the FKS model has both hard and soft components for all sizes,
but in smoothly varying amounts.}. For
dipoles whose transverse size is less than about 0.8~fm, the resulting
behaviour is not unlike the FKS dipole cross-section at $W = 75$~GeV, which 
is the mean energy of the data, and the model gives a good account of 
H1 data in this region. 
However at higher energies the model develops a rather
artificial dependence on the dipole size: as can be seen from the dotted 
curve in Fig. \ref{fig:dd_dipole} for $W=300$~GeV, the dipole cross section 
not only does not increase monotonically with dipole size, but develops 
a large discontinuity at the matching point ($d_{\perp} \approx 0.3$~fm).
The other obvious difference is that the cross-section is much larger 
than the FKS cross-section at large $\dperp$. However for DIS this can be 
compensated by differences in the photon wave-function, which is uncertain in this
region, as noted at the end of section 2.1. FKS resolved this ambiguity by
considering the DDIS reaction of eq.(\ref{DDIS}), which is more 
sensitive 
to large dipoles and involves a strikingly different combination of
wavefunction and cross-section. It would be interesting to see the predictions
of the Donnachie-Dosch model for this reaction.

\subsection{The MFGS model}

\noindent This model \cite{MFGS1,MFGS2} is directly based on the known 
behaviour of 
hard small-$x$ QCD processes, i.e. that they are driven by the gluon 
distribution at small $x$. 
Using the phenomena of colour transparency, it directly relates the dipole 
cross-section at small $\dperp$ to leading order (LO) gluon distributions at
large $Q^2$. 
To leading order in $\ln Q^2$, and within the small $x$ limit, 
the total photon-proton cross-sections are given by expressions of the form eq.(\ref{absorptive}) with 
a QED  wavefunction and a dipole cross-section\footnote{This formula is
implicit in most perturbative two gluon models. For an explicit derivation see
\cite{derivation}.} 
\begin{equation}
\hat{\sigma}_{\mbox{{\tiny pQCD}}}(x,\dperp) =\frac{\pi^{2} \dperp^{2}}{3}\alpha_{s}(\bar{Q}^{2})
xg(x^{\prime},\bar{Q}^2) \, ,
\label{MFGS1}
\end{equation}
where $xg(x^{\prime},\bar{Q}^2)$ is the LO gluon distribution of the proton. 
At leading log it is sufficient to choose $x' = x$  and ${\bar Q}^2 = Q^2$. 
However in the MFGS model (specified fully in \cite{MFGS1}) an attempt to go
beyond leading log was made by introducing $\dperp$-dependence into 
the scales, $x^{\prime}$ and $\bar{Q}^2$. 
For the four momentum scale $\bar{Q^{2}}$, the phenomenological relation
\begin{equation}
\bar{Q^{2}}=\frac{\lambda}{\dperp^{2}}
\label{MFGS2}
\end{equation}
was assumed where $\lambda = <\!\dperp^2\!> Q^2$. A theoretical procedure for defining 
$<\!\dperp^2\!>$ using the integral in $\dperp$ for $F_L$ in \cite{FKopfS} gave a value 
of $\lambda \approx 10$. This value was used in \cite{MFGS1}, but it was later discovered 
that the inclusive cross sections are rather insensitive to its precise value in the range 
$\lambda = 4-15$ and that the lower value of $\lambda = 4$ appears to be favoured by the 
$J/\psi$-photoproduction data \cite{MFGS2}. We adopt this lower value in what follows.
The momentum fraction required to create a quark-antiquark pair of mass  
$M^2_{q {\bar q}} = (k_{T}^2 + m_q^2)/(z(1-z))$ is
\begin{equation}
x^{\prime}=\frac{M_{q {\bar q}}^2+Q^2}{Q^{2}+W^{2}} \, ;
\end{equation}
since $k_{T}$ is Fourier conjugate to $\dperp$ the following relationship was
adopted \cite{MFGS1}
for a dipole of given transverse size $\dperp$:
\begin{equation}
x^{\prime}= x \left[1 + 0.75 \frac{\lambda}{\dperp^2 (Q^2 + 4 m_q^2)} \right] \, .
\end{equation}
The dipole cross-section was then evaluated using the CTEQ4L gluon distributions \cite{cteq4l} 
for a region $ \dperp \le d_{T, c}$ in which eq.(\ref{MFGS1}) is 
appropriate, where the boundary  $d_{T, c}$ is specified below. 
For large dipoles $ \dperp \ge d_{T, \pi} = 0.65$ fm, the form
\begin{equation}
\hat{\sigma} (\dperp > d_{T, \pi}) =  \hat{\sigma} (\pi P) \frac{3 \, \dperp^2}
{2 \, \dperp^2 + d_{T, \pi}^2}  \, \left( \frac{x_0}{x} \right)^{0.08}    \; ,
\end{equation}
with an $x$-dependence characteristic of soft Pomeron exchange, is used, where $x_0 = 0.01$ and
 the value at $\dperp = d_{T, \pi}\, , \; x = x_0$ is matched to the pion-proton total cross-section,
$\sigma(\pi P) = 24$~mb. In the intermediate region
 $d_{T, c} < \dperp < d_{T, \pi}$, the dipole cross-section
was linearly interpolated between the boundary values at $d_{T, c}$ and $d_{T, \pi}$.  

\begin{figure}[htbp]
  \begin{center}
   \includegraphics[width=10cm,height=6cm]{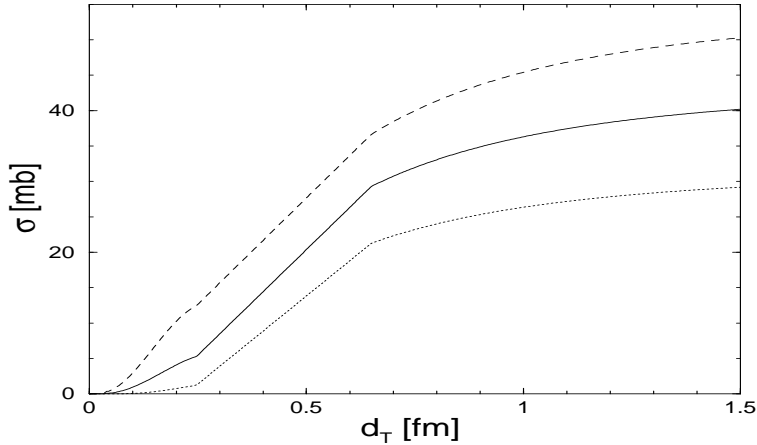}      
    \caption{The MFGS dipole at $W = 10, 75, 300$~GeV (dotted, solid and dashed lines, respectively) 
at fixed $Q^2 = 1$~GeV$^2$ , corresponding to $x \approx 10^{-2}, 2. 10^{-4}$ and $10^{-5}$, respectively.}
    \label{fig:mfgsdipole}
  \end{center}
\end{figure}
For moderate $x$, the point $d_{T, c}$ is set by the boundary of the perturbative region in $\dperp$: $d_{T, c} = d_{T, 0} \equiv \sqrt{\lambda/Q_0^2} = 0.246$~fm, for $Q_0 = 1.6$~GeV which is the starting scale for the CTEQ4L partons (below this scale they are not defined). 
The resulting dipole cross-section for $Q^2 = 1$~GeV$^2$ is shown in Fig. \ref{fig:mfgsdipole} for 
various values of $W$. 

At small enough $x$, as a result of the rising gluon density, the small dipole cross section increases faster than the large dipole one and soon reaches hadronic size (tens of mb). This threatens to spoil the monotonic increase of ${\hat \sigma}$ with $\dperp$. To prevent this the MFGS model
implements taming corrections that guarantee that the small dipole cross
section cannot reach more than half its value at $d_{T, \pi}$. 
This constraint implies a $d_{T, c}$ that shifts to increasingly small $\dperp$. 
This correction is not crucial in the HERA region for $\lambda=4$, 
but does becomes important above it.

The parameters of the model are not adjusted to fit data, but nonetheless  
good semi-quantitative accounts of the deep inelastic
scattering \cite{MFGS1}  and $J/\psi$ photoproduction data \cite{MFGS2} were 
obtained. For exclusive diffractive processes, such as vector meson production or DVCS, 
it is necessary to include GPDs, parameterized in terms of skewedness, $\delta$, $x'$ and 
$\bar{Q^2}$, rather than the ordinary ones used in eq.(\ref{MFGS1}). For DVCS, 
$\delta = x_{Bj}= Q^2/W^{2}$. To implement leading order GPDs, 
we adapted the skewed evolution package developed by Freund and Guzey \cite{freund1},  
using the CTEQ4L gluon distributions \cite{cteq4l} as input to the LO skewed evolution.

The model focuses on small dipoles, and while the behaviour for 
large dipoles is a sensible guess, no detailed attempt to resolve the 
intrinsic ambiguity in the wavefunction and cross-section
for large dipoles discussed earlier has been made. In what follows we shall 
restrict this model to $Q^2 > 1$ GeV$^2$, where one is relatively insensitive
to this region due to the smallness of the wavefunction.

\section{Virtual Compton scattering}

\label{sec:dvcsfr}

In the colour dipole model, virtual Compton scattering is again given  
by Fig. \ref{fig:fluct}, but with a real photon in the final state,
leading to
\begin{equation}
Im~{\cal A}^{\mbox{{\tiny DVCS}}} (W^2, Q^2, t=0) = \int dz ~\mbox{d}^2 \dperp ~\psi^*_{T}(z,\dperp,Q^2)  ~{\hat \sigma} (s^*, z, \dperp) ~\psi_{T} (z,\dperp,0) \, ,
\label{DVCS1}
\end{equation}
for the imaginary part of the DVCS amplitude at zero momentum transfer. 
Thus, our dipole models yield no-free-parameter predictions for this process.
In this section we compare the predicted behaviours of the amplitude in the two
models, leaving the comparison between predictions and experiment to section \ref{sec:res}.

\begin{figure}[htbp]
  \begin{center}
   \includegraphics[width=10cm,height=6cm]{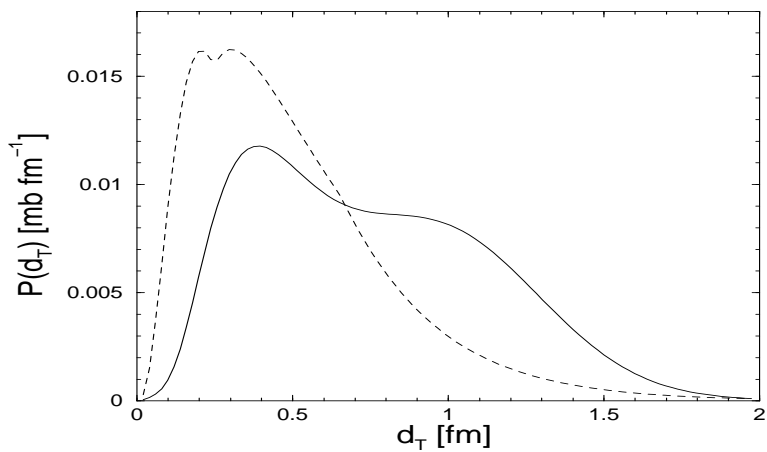}
    \caption{Profile in transverse dipole size for $Q^2 = 1$~GeV$^2$ and $W = 75$~GeV, 
employing the FKS (solid line) and MFGS (dashed line) models for the dipole cross section.}
    \label{profile1.eps}
  \end{center}
\end{figure}

\begin{figure}[htbp]
  \begin{center}
   \includegraphics[width=10cm,height=6cm]{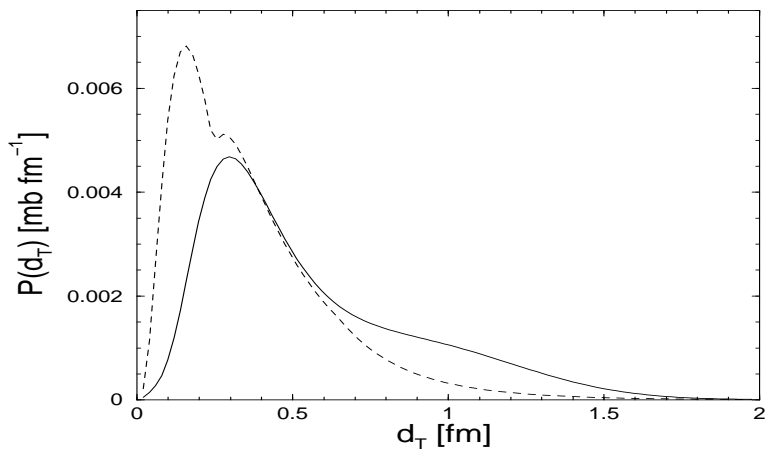}
    \caption{Profile in transverse dipole size for $Q^2 = 10$~GeV$^2$ and $W = 75$~GeV, 
employing the FKS (solid line) and MFGS (dashed line) models for the dipole cross section.}
    \label{profile10.eps}
  \end{center}
\end{figure}

We start by comparing the contributions to the amplitude arising from dipoles
of different size. To do this we perform 
the angular and 
$z$ integrations to rewrite eq.(\ref{DVCS1}) in the form 
\begin{equation}
Im~{\cal A} (W^2, Q^2, t=0) = 2 \pi  \int  {\mbox d} \dperp ~p(\dperp,s,Q^2) \, ,
\label{profile}
\end{equation}
where the profile function 
\begin{equation}
p (\dperp, s^*, Q^2) = \int dz~\dperp~\psi^*_{T}(z,\dperp,Q^2)~{\hat \sigma} (s^*, z, \dperp)~\psi_{T} (z,\dperp,0)  \, ,
\label{DVCSprofile}
\end{equation}
gives 
the relative contributions arising from dipoles of different size $\dperp$.
The results are  shown for both models at the mean energy, $W=75$~GeV, of
the H1 data  in Figs. \ref{profile1.eps}, \ref{profile10.eps}.  As  $Q^2$ increases, 
the profile shifts to smaller $\dperp$.
The FKS model has a larger contribution from large dipoles than the 
MFGS model,  
although the forward amplitudes, obtained by integrating over
all transverse sizes, are similar over a wide range of $W$ and $Q^2$, 
as we shall see.  

\begin{figure}[htbp]
  \begin{center}
   \includegraphics[width=10cm,height=6cm]{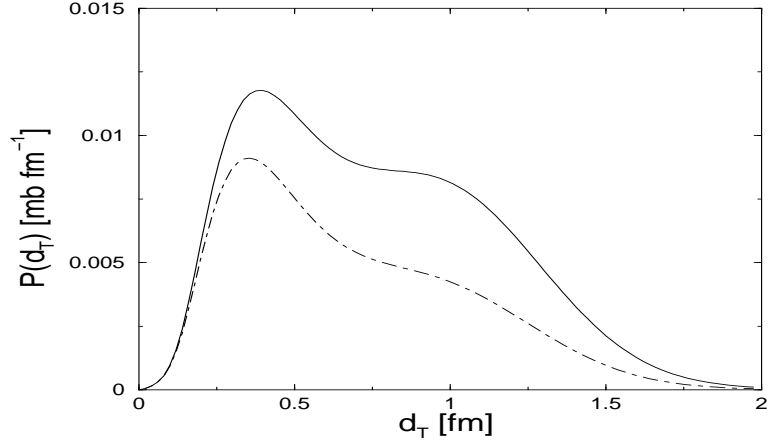}
    \caption{Comparison of the profile functions of eqs.(\ref{DVCSprofile},
\ref{DISprofile}) for DVCS (solid line) and  transverse DIS (dot-dash line) 
respectively 
 at $Q^2 = 1$~GeV$^2$ and $W = 75$~GeV, 
employing the FKS model for the dipole cross section.}
    \label{compare1.eps}
  \end{center}
\end{figure}

\begin{figure}[htbp]
  \begin{center}
   \includegraphics[width=10cm,height=6cm]{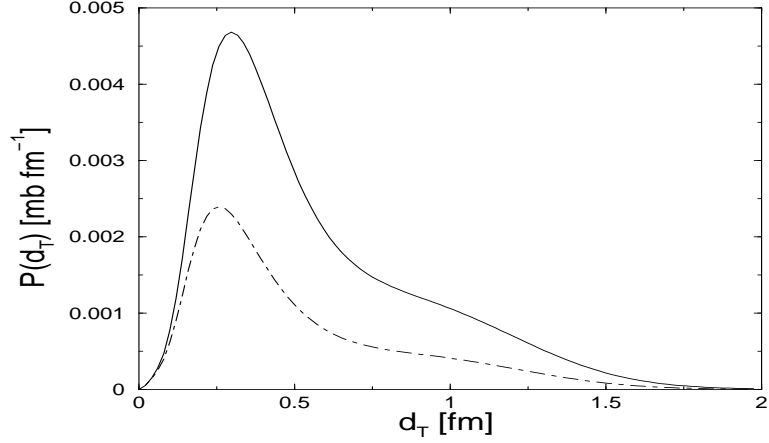}
    \caption{Comparison of the profile functions of eqs.(\ref{DVCSprofile},
\ref{DISprofile}) for DVCS (solid line) and 
for transverse DIS (dot-dash line) at $Q^2 = 10$~GeV$^2$ and $W = 75$~GeV, 
employing the FKS model for the dipole cross section. }
    \label{compare10.eps}
  \end{center}
\end{figure}

It is also interesting to compare eq.(\ref{DVCSprofile}) to the corresponding
profile function
\begin{equation}
\tilde{p}_T (\dperp, s^*, Q^2) = \int dz~\dperp~\psi^*_{T}(z,\dperp,Q^2)~{\hat \sigma} (s^*, z, \dperp)~\psi_{T} (z,\dperp,Q^2)  \, ,
\label{DISprofile}
\end{equation}
for the  forward Compton scattering amplitude in which  both 
transverse photons have the same $Q^2$, which is
related by the optical theorem to the transverse cross-section in DIS.
The characteristic behaviour differences observed between DIS and DVCS are 
illustrated in Figs. \ref{compare1.eps} and \ref{compare10.eps}, 
using the FKS dipole model. For $Q^2 = 0$ the two profiles are obviously identical,
and for all $Q^2$  they  become identical for small $\dperp$, 
 since to leading order in $1/\dperp$, the transverse wavefunction
\begin{eqnarray}
  \label{eq:smallr}
  |\psi_{T}^0(z,\dperp, Q^2)|^{2} & \rightarrow & 
\frac{3}{2 \pi^{2}}\alpha_{e.m.} \sum_{q=1}^{n_{f}}e_{q}^
{2} \frac{z^{2} + (1-z)^{2}}{ \dperp^2}  
\end{eqnarray}
independent of $Q^2$.
However, as can be seen, when $Q^2$ increases the large dipoles are 
less suppressed in 
the DVCS case than in the DIS case, so that the former is the ``softer'' process at any given $Q^2$.

\begin{figure}[htbp]
  \begin{center}
   \includegraphics[width=10cm,height=6cm]{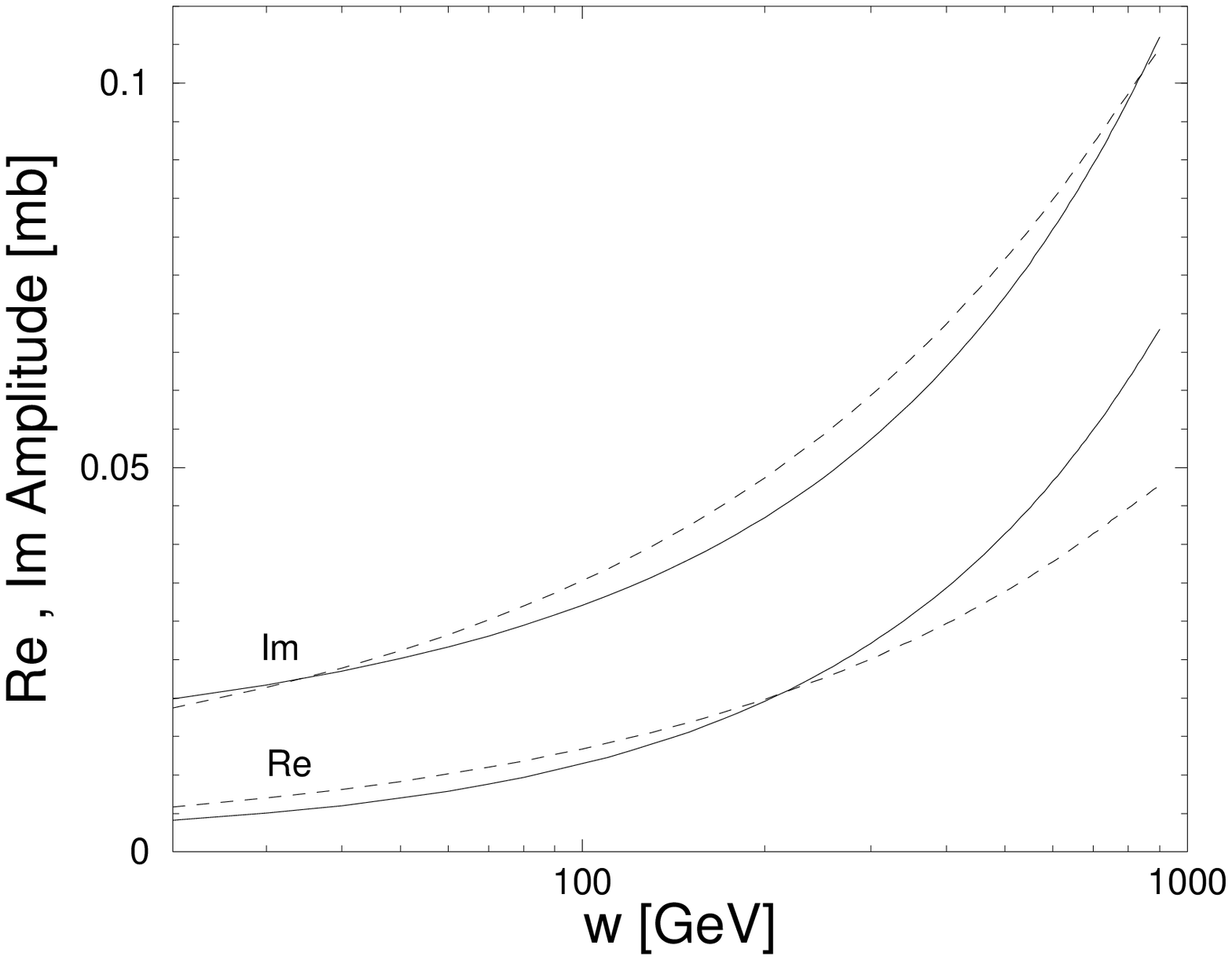}
    \caption{Real and imaginary parts of the DVCS amplitude for the FKS (solid lines) and MFGS (dashed lines) dipole models for $Q^2= 4.5$~GeV$^2$.}
    \label{fig:reim}
  \end{center}
\end{figure}

\begin{figure}[htbp]
  \begin{center}
   \includegraphics[width=10cm,height=6cm]{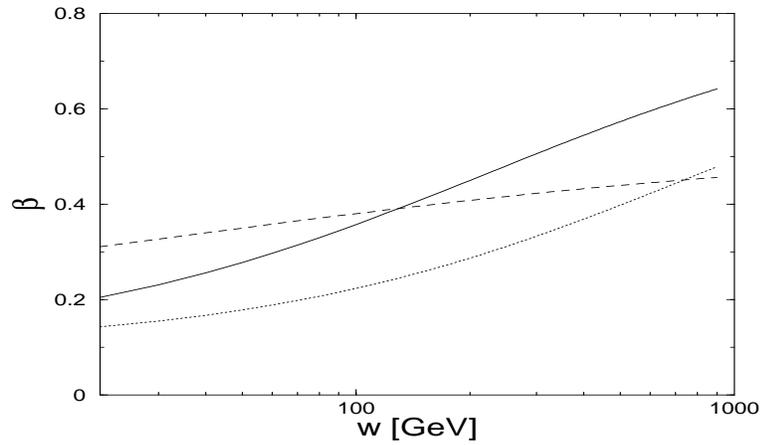}      
    \caption{The ratio, $\beta$, of the real to imaginary parts of the DVCS 
forward amplitude at $Q^2= 4.5$~GeV$^2$: FKS (solid 
line); MFGS (dashed line); and FKS at $Q^2 = 0$~GeV$^2$ (dotted line).}
    \label{fig:betalogw}
  \end{center}
\end{figure}

Returning to DVCS, the imaginary part of the amplitude is trivially 
obtained by integrating eq.(\ref{profile}) over all $\dperp$.  In the 
FKS model we have a sum of two Regge contributions, and the real part 
is easily computed from the corresponding signature factors; for the 
MFGS model, results were obtained using dispersion relations, as in 
\cite{MFGS2}. The results for real and imaginary parts, using both models, 
are plotted as a function of $W$ in Fig. \ref{fig:reim} at the mean $Q^2$ 
of the H1 data, and the ratio is plotted in Fig. \ref{fig:betalogw}. 
We extend the energy range out to the THERA  range (see e.g. \cite{thera}) 
and one can clearly see that the FKS model has a steeper energy dependence 
at very high energies.

\begin{figure}[htbp]
  \begin{center}
   \includegraphics[width=10cm,height=6cm]{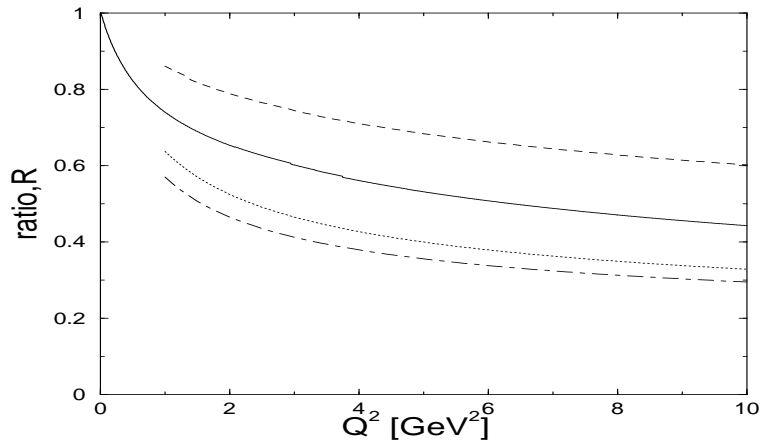}      
    \caption{The ratio of the forward amplitudes for DIS and 
DVCS (cf. eq.(\ref{ratio1})) at $W = 75$~GeV: FKS (solid line); MFGS (dashed line); and the prediction
of eq.(\ref{ratio2}) for  $M_0^2=0.6$~GeV$^2$ (dotted line) and
$M_0^2=0.4$~GeV$^2$ (dot-dashed line), respectively.}
    \label{fig:ratio}
  \end{center}
\end{figure}

Finally, in our introduction we noted that in \cite{FFS}, the ratio 
\begin{equation}
R \equiv \frac{Im~{\cal A}(\gamma^* N \rightarrow \gamma^* N )_{t=0}}
{Im~{\cal A}(\gamma^* N \rightarrow \gamma  N )_{t=0}}
\label{ratio1}
\end{equation}
of the imaginary parts of the forward amplitudes for DIS and DVCS   
was estimated at the input $Q^2 = 2.6$~GeV$^2$ using a simple aligned 
jet model, in order to infer the 
input generalized parton distributions for QCD evolution. 
Explicitly, this model  gives \cite{FFS} 
\begin{equation}
R = \frac{Q^2}{Q^2 + M_0^2} \, \ln^{-1}(1 + Q^2 / M_0^2)
\label{ratio2}
\end{equation}
where $M_0^2$ is estimated to be in the range 0.4 - 0.6~GeV$^2$. The 
predictions of our models are compared with eq.(\ref{ratio2}) in
 Fig. \ref{fig:ratio}, suggesting somewhat larger values
at  $Q^2 = 2.6$~GeV$^2$, as can be seen.

\section{The lepton level process}

\label{bh}

\subsection{Definition of the DVCS cross section}

Virtual Compton scattering is accessed experimentally through the leptonic 
process: 
\begin{equation}
e^{\pm} (k) + p(p) \rightarrow e^{\pm} (k') + p (p') + \gamma (q')
\end{equation}
\noindent where the four momenta of the incoming and outgoing particle are
given in brackets. As well as DVCS, the Bethe-Heitler process (BH), in which 
the photon is radiated by the initial or final state lepton, also contributes. 
On integrating over the azimuthal angle (defined below), the
interference term between the two processes vanishes in the limit of large $Q^2$, 
and the differential cross-section can be written as:
$$
\frac{d^{2}\sigma}{dy dQ^{2}} = 
\frac{d^{2}\sigma^{DVCS}}{dy dQ^{2}}
+ \frac{d^{2}\sigma^{BH}}{dy dQ^{2}} \, , 
$$ 
where $y \equiv (k-k') \cdot P/ (k \cdot P)$ and $Q^2 = - (k-k')^2$. 
For DVCS, $Q^2$ is the magnitude of the virtuality of the (spacelike) virtual 
photon and, in the proton's rest frame, $y$ 
is the fraction of the incoming electron energy carried by the virtual photon. 
Neglecting the lepton and proton masses we have $y \approx (Q^2 + W^2)/S$, where $S = (k+p)^2$ is the square of the lepton-proton centre-of-mass energy. 
The Bethe-Heitler contribution is essentially known in terms 
of the Dirac and Pauli form factors (see e.g. eqs.(18, 27) of \cite{bmns2}) and
can be easily calculated and subtracted from the total to leave
\begin{equation}
\frac{d^{2} \sigma^{DVCS}} {dy dQ^{2}}  =\frac{\alpha_{e.m.}}{2 \pi Q^{2} y}
~[1+(1-y)^{2}] ~\sigma(\gamma^{*}p\rightarrow\gamma p) \, . 
\end{equation}
\noindent Making a trivial change of variable from $y$ to 
$W = \sqrt{(k - k' + P)^2}$ yields:
\begin{equation}
\frac{d^{2}\sigma^{DVCS}}{dW dQ^{2}}  =\frac{\alpha_{e.m.}}{\pi Q^{2} W}
~[1+(1-y)^{2}] ~\sigma(\gamma^{*}p\rightarrow\gamma p) \, . 
\end{equation}
\noindent We now have a convenient form for comparing to the data on 
$\sigma(\gamma^* P \rightarrow \gamma p)$ which is binned in $Q^2$ and $W$.
Assuming the usual exponential dependence in $t = (p-p')^2 <0$, i.e. $\rm{e}^{Bt}$, the total $\gamma^{*} - p$ cross section is given by:  
\begin{equation}
\sigma (\gamma^{*} p \rightarrow \gamma p)= \frac{1}{B}  ~\frac{d \sigma}{dt} \mid_{t=0} \, ,
\label{sigtot}
\end{equation}
where, with our definition for ${\cal A}^{DVCS}$, 
\begin{equation}
\frac{d\sigma}{dt} \mid_{t=0}
 = \frac{(\rm{Im} {\cal A})^2}{16 \pi}\, (1 + \beta^2)
\label{dcs} 
\end{equation} 
and $\beta$ is the ratio of the real to the imaginary part of the forward 
virtual Compton scattering amplitude  ${\cal A}$ at $t = 0$.

\subsection{Definitions of asymmetries}

A unique and attractive feature of DVCS is the interference with the 
Bethe-Heitler process which offers the rare chance to isolate both real 
and imaginary parts of the diffractive amplitude via azimuthal angle 
asymmetries \cite{bmns2}. These asymmetries are conveniently discussed  
in a special frame \cite{bkms} with the proton at rest such that the direction of the vector $q \equiv k-k'$ defines the negative $z$-axis.  Without loss of generality we can choose the incoming 
electron to have only a non-zero component along the positive $x$-axis in the transverse ($x-y$) plane. In this frame we have the following four-vectors:
\begin{eqnarray}
&k = (k_0, k_0 \sin \theta_e, 0 , k_0 \cos\theta_e), \quad q = (q_0, 0, 0, -|q_3|) \\
& P = (M, 0, 0, 0) , \quad P' = (P'_0, |{\bf P'}| \cos\phi \sin \theta_H, |{\bf P'}| \sin\phi \sin\theta_H, |{\bf P'}| \cos\theta_H) ,
\end{eqnarray}
where the angle of interest, $\phi$, is the azimuthal angle between the lepton ($x-z$) and hadron scattering planes. 

The motivation for using this frame is that the frame-dependent expression  
for the $u-$channel BH lepton propagator has a particularly simple Fourier 
expansion in the angle $\phi$. In \cite{bmns2} a slightly different frame is used and an explicit expression for the $u-$channel propagator is given, up to terms of order ${\cal O} (1/Q^3)$ (cf. eq.(21) of \cite{bmns2}): 
\begin{equation}
(k-q^{'})^2 = - \frac{(1-y) Q^2}{y} (A_0 + A_1 \cos \phi + A_2 \cos 2 \phi + \cdots ) \, ,
\label{fourier}
\end{equation}
\noindent where
\begin{eqnarray}
A_0 & = & 1 - \frac{t}{Q^2} \left(\frac{1}{2} + \frac{(1-x) (1 - 2~t_{\mbox{min}}/t )}{(1-y)}  \right) \, , \nonumber \\
A_1 & = & 2 \sqrt{\frac{-t}{Q^2}} \sqrt{\frac{(1 - t_{\mbox{min}}/t) (1-x)}{(1-y)}} \, . \label{fourierc}
\end{eqnarray}
\noindent In the frame used here \cite{bkms} the Fourier series terminates at
$\cos \phi$ ($A_n = 0, n \geq 2$). We explicitly include factors of $1/(A_0 +
A_1 \cos \phi)$ as appropriate in our numerical results. We use a code written
for \cite{afmm2} which approximately implements eq.(18) and eqs.(24,27,30) of
\cite{bmns2} (the latter neglect terms of ${\cal O}(1/Q)$, i.e. they use $A_0 =
1$ and $A_1$ = 0). The code includes the above expansion of the $u-$ channel
BH lepton propagator in our frame (taking its full $\phi$ and $y$-dependence 
into account, up to corrections of ${\cal O}(1/Q^3)$). 
The asymmetries of interest involve the quadruple differential cross section on the lepton level 
$$
d \sigma^{DVCS+BH} =\frac{d \sigma^{(4)} (ep \rightarrow ep\gamma)}{dx dQ^2 dt d\phi} \, .
$$ 
In order to proceed it was necessary to convert our amplitude (cf. eq.(\ref{DVCS1})) to the dimensionless unpolarized helicity non-flip amplitude, ${\cal H}_1$, appearing in eqs.(24,27,30) of \cite{bmns2}, since at small $x$ and moderate $t$ the contributions of the polarized, ${\tilde  {\cal H}}_1, {\tilde {\cal E}}_1$, and unpolarized helicity-flip, ${\cal E}_1$, DVCS amplitudes are negligible. Assuming a simple exponential $t-$dependence on the amplitude level the conversion factor is:
\begin{equation} 
{\cal H}_1  = \exp (Bt/2) \frac{W^2}{4 \pi \alpha_{e.m.} \hc } A^{\mbox{dipole}} 
\end{equation} 
\noindent where  $W$ is in units of GeV, so the standard conversion factor 
$\hc = 0.389$~GeV$^2$~mb is necessary to make ${\cal H}_1$ dimensionless.

Using the special frame defined above to specify the 
azimuthal angle, the asymmetries are defined as follows (see also \cite{afmm2}):

\begin{itemize} 
\item The (unpolarized) azimuthal angle asymmetry (AAA), measured in 
the scattering of an unpolarized probe on an unpolarized target, is 
defined by 
\begin{center} 
\bea 
\mbox{AAA} =\frac{\Big.\int^{\pi/2}_{-\pi/2} d\phi 
(d\sigma^{DVCS+BH}-d\sigma^{BH}) - \Big.\int^{3\pi/2}_{\pi/2} d\phi (d\sigma^{DVCS+BH}-d\sigma^{BH})}{\Big.\int^{2\pi}_{0} d\phi (d\sigma^{DVCS+BH}-d\sigma^{BH})} \,  
\label{aaadef} 
\eea 
\end{center} 
where $d \sigma^{BH}$ is the pure BH term. The above approximation for the
$u-$channel BH propagator leads to a non-trivial $\phi-$dependence of the pure
BH term. To directly access the DVCS amplitudes, via the interference term, we
define AAA with this piece subtracted. Note that with this ``subtracted  
definition'' the magnitude of AAA may become greater than unity in certain regions. 
\item The single spin asymmetry (SSA), measured in the scattering of a 
longitudinally polarized probe on an unpolarized target, is defined by 
\begin{align} 
&\mbox{SSA} = \frac{\int^{\pi}_{0} d\phi \Delta\sigma^{DVCS+BH} - \int^{2\pi}_{\pi} d\phi \Delta\sigma^{DVCS+BH}}{\int^{2\pi}_{0} d\phi (d\sigma^{DVCS+BH,\uparrow}+d\sigma^{DVCS+BH,\downarrow})} \, , 
\label{ssadef} 
\end{align} 
where $\Delta\sigma = d\sigma^{\uparrow} - d\sigma^{\downarrow}$ and 
$\uparrow$ and $\downarrow$ signify that the lepton is polarized along or against its direction of motion, respectively. 
\item The charge asymmetry (CA) in the scattering of an unpolarized probe 
on an unpolarized target: 
\begin{align} 
&\mbox{CA} =\frac{\Big.\int^{\pi/2}_{-\pi/2} d\phi 
\Delta d^C\sigma^{DVCS+BH} - \Big.\int^{3\pi/2}_{\pi/2} d\phi \Delta d^C\sigma^{DVCS+BH}}{\Big.\int^{2\pi}_{0} d\phi (d^+\sigma^{DVCS+BH}+d^-\sigma^{DVCS+BH})} \, , 
\label{cadef} 
\end{align} 
where $\Delta d^C\sigma = d^{+}\sigma - d^{-}\sigma$ corresponds to the difference 
of the scattering with a positron probe and an electron probe. 
\end{itemize} 

In the small-$x$ limit, required for the dipole approximation, and at large $Q^2$, AAA and CA are directly proportional to the real part of the DVCS amplitude, and SSA to the imaginary part of the unpolarised amplitude (cf. eq.(30,40,43) of \cite{bmns2})\footnote{At large $x$ one also gets a contribution from the polarised and helicity-flip DVCS amplitudes, $ {\tilde {\cal H}}_1, {\cal E}_1, {\tilde {\cal E}}_1 $, which are negligible in the small $x$ region (cf. eqs.(24, 30) of \cite{bmns2}).}. 

\begin{figure}[htbp]
  \begin{center}
   \includegraphics[width=10cm,height=6cm]{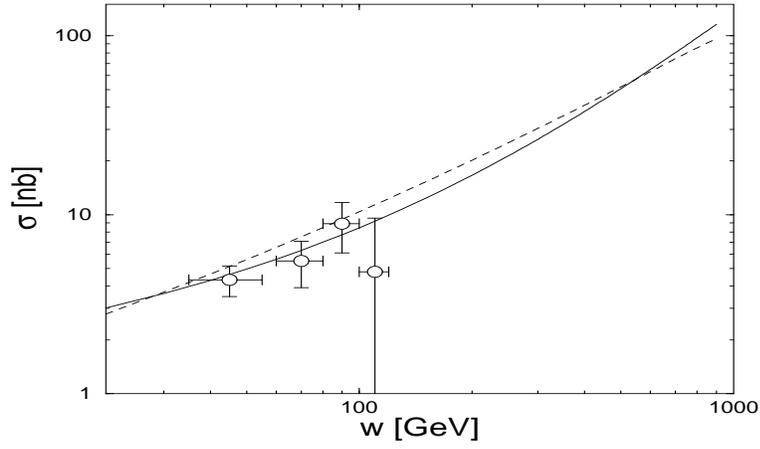}
    \caption{The energy dependence of the photon level DVCS cross section at fixed $Q^2 = 4.5$~GeV$^2$: FKS (solid line); MFGS (dashed line).}
    \label{fig:photq}
  \end{center}
\end{figure}

\begin{figure}[htbp]
  \begin{center}
   \includegraphics[width=10cm,height=6cm]{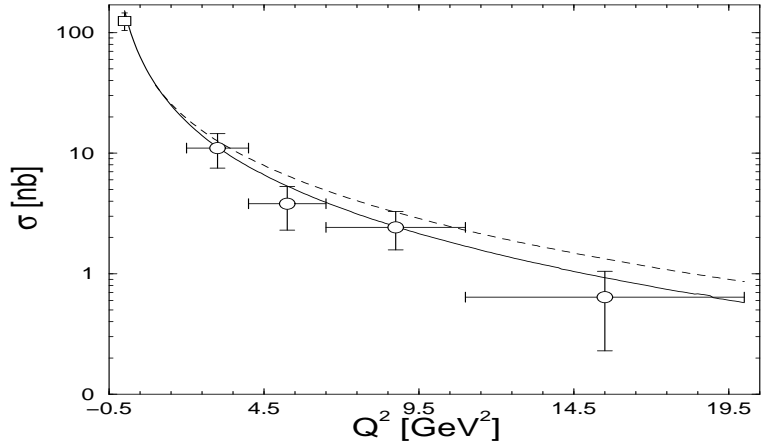}
    \caption{The $Q^2$-dependence of the photon level DVCS cross section at fixed $W = 75$~GeV: 
FKS (solid line); MFGS (dashed line).}
    \label{fig:photw}
  \end{center}
\end{figure}

\begin{figure}  
\centering 
\includegraphics[width=10cm,height=7cm]{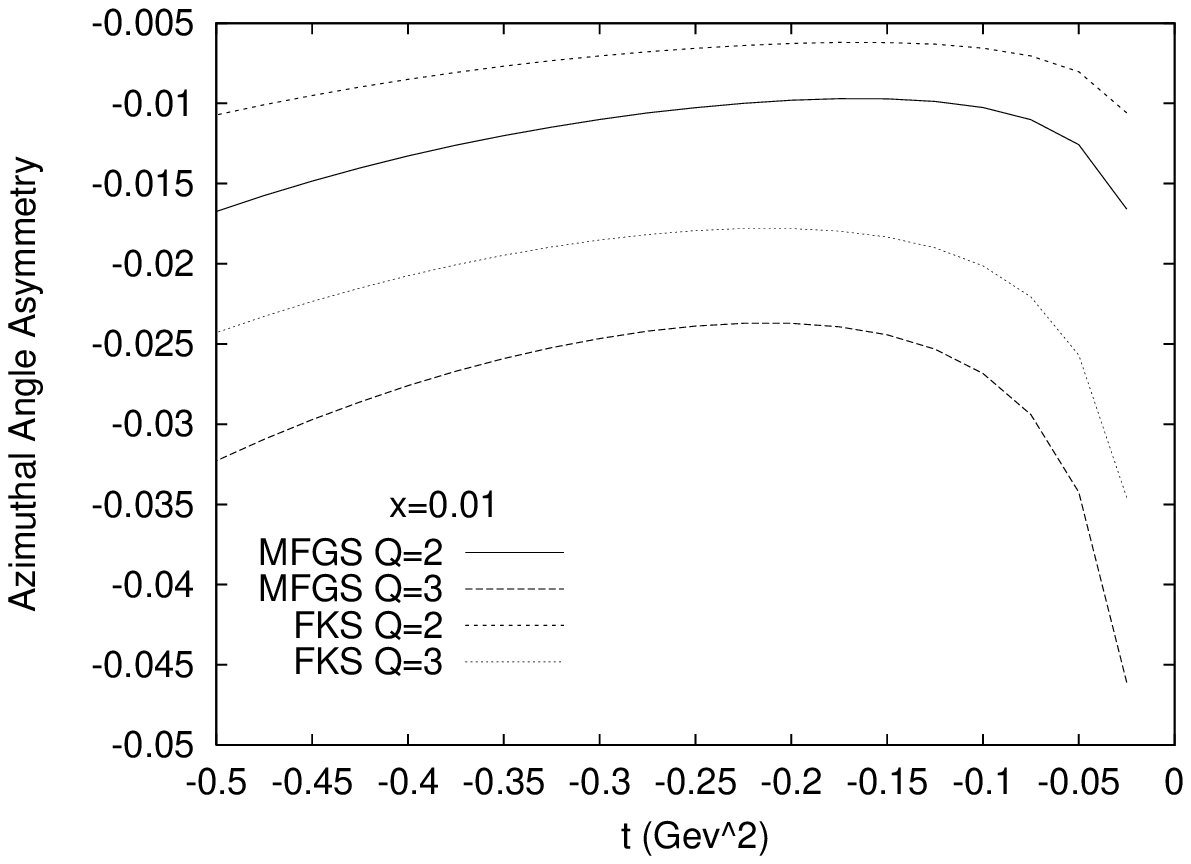}    
\includegraphics[width=10cm,height=7cm]{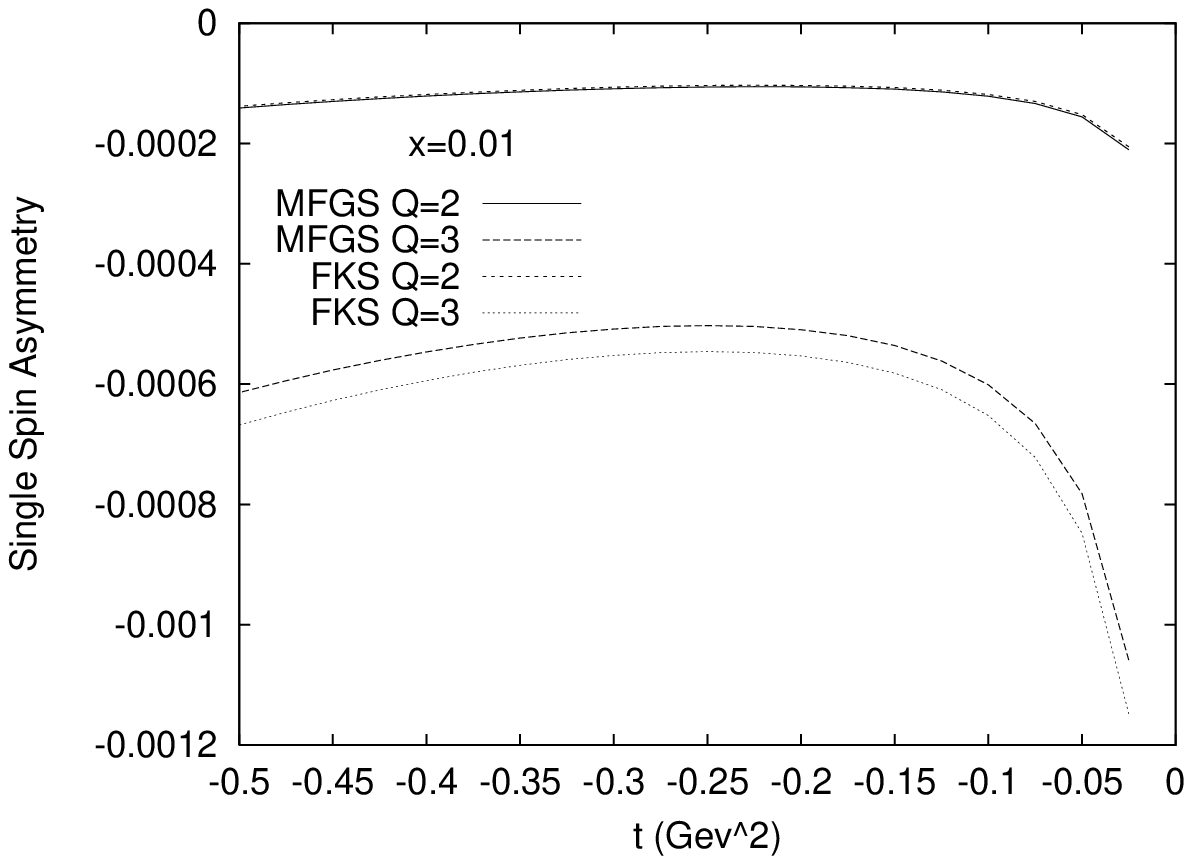}    
\includegraphics[width=10cm,height=7cm]{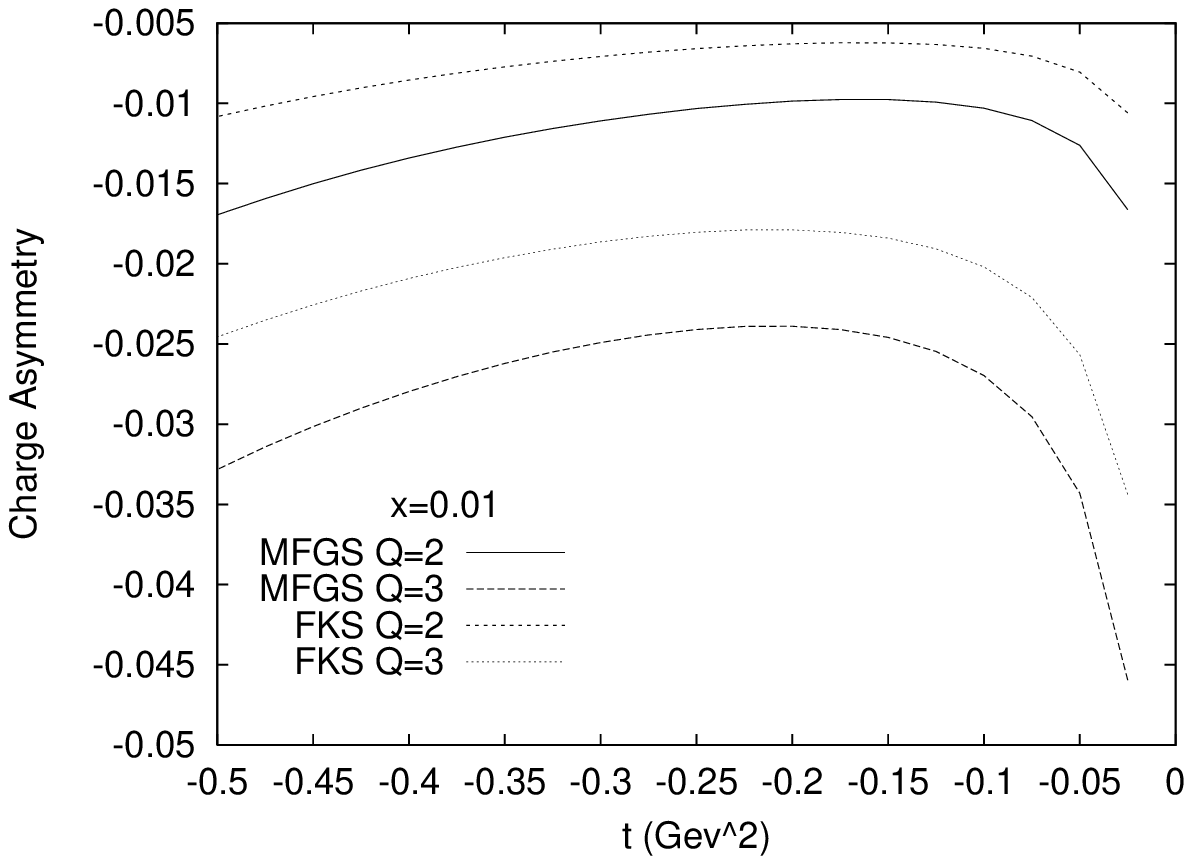}    
\caption{Compendium of results on asymmetries for fixed $x=10^{-2}$, at two values of $Q=2,3$~GeV, accessible in the HERA kinematic range.}
\label{asymx2}
\end{figure}

\begin{figure}  
\centering 
\includegraphics[width=10cm,height=7cm]{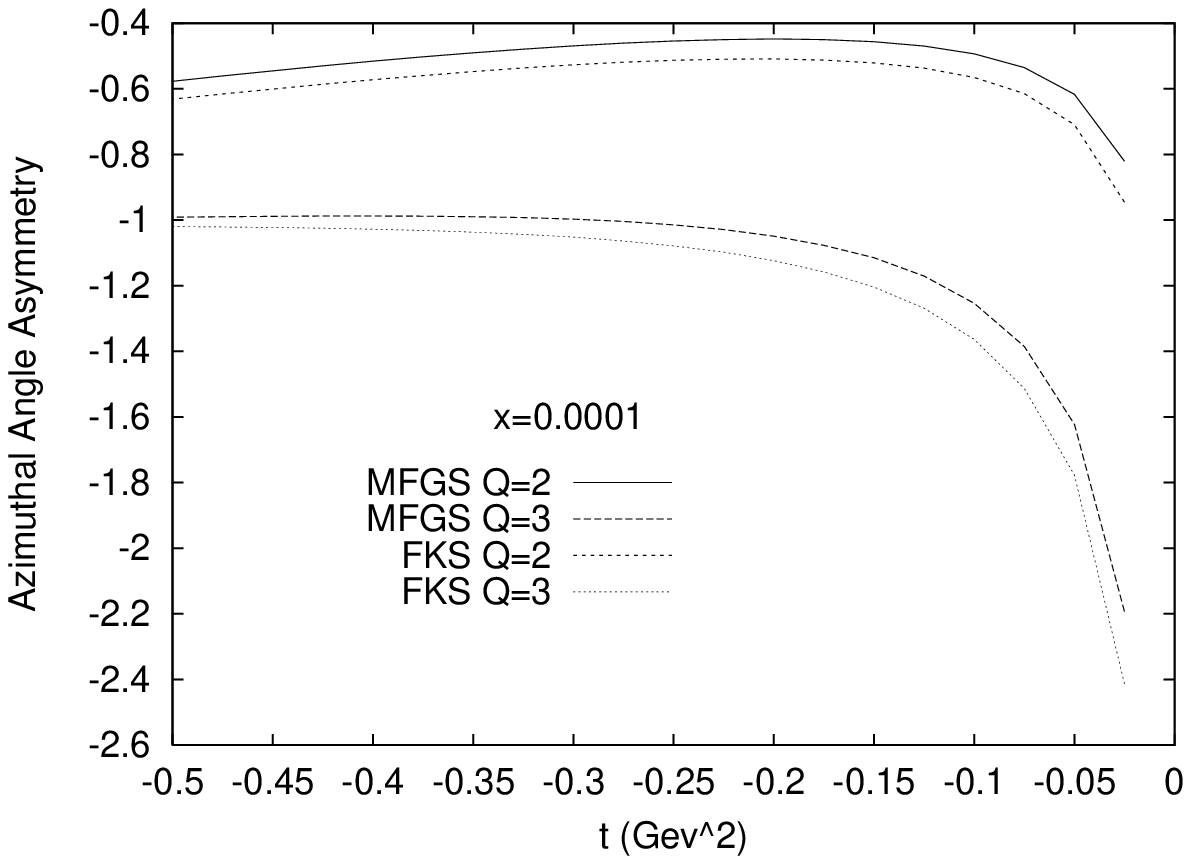}    
\includegraphics[width=10cm,height=7cm]{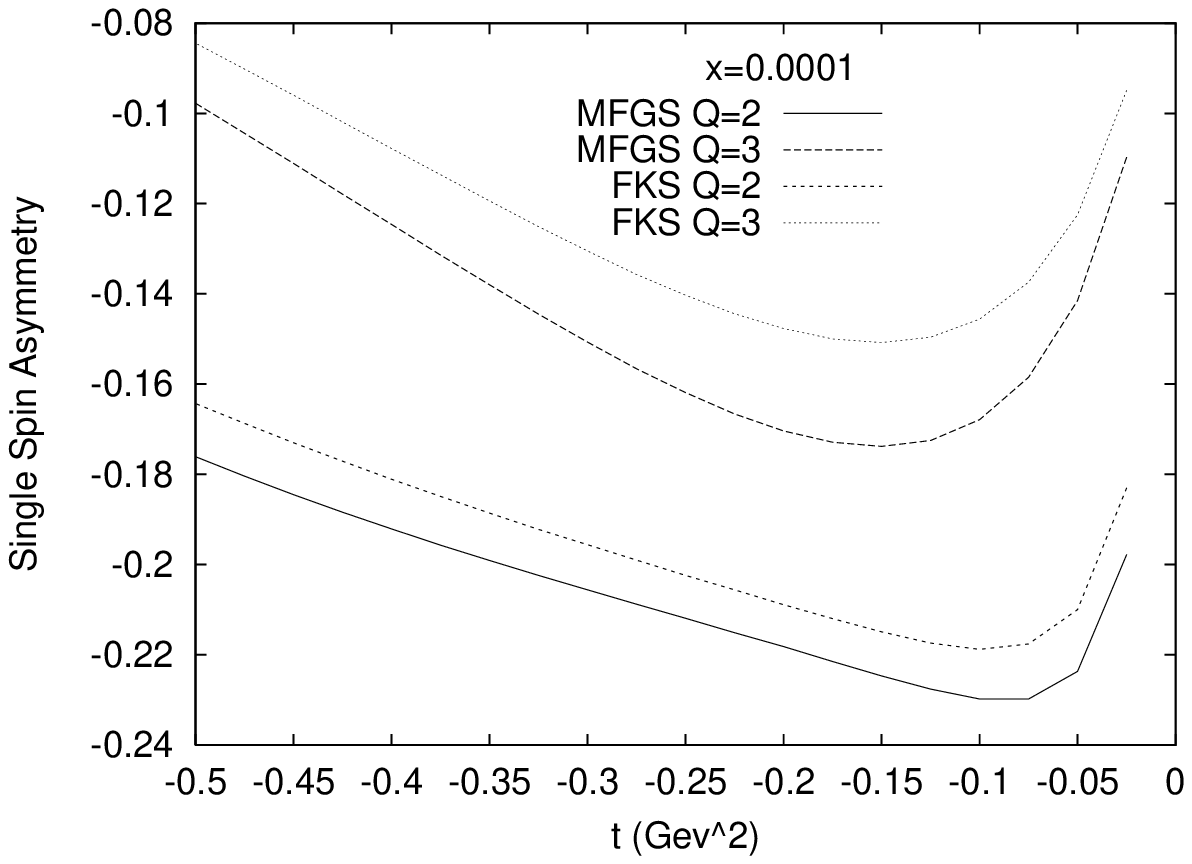}    
\includegraphics[width=10cm,height=7cm]{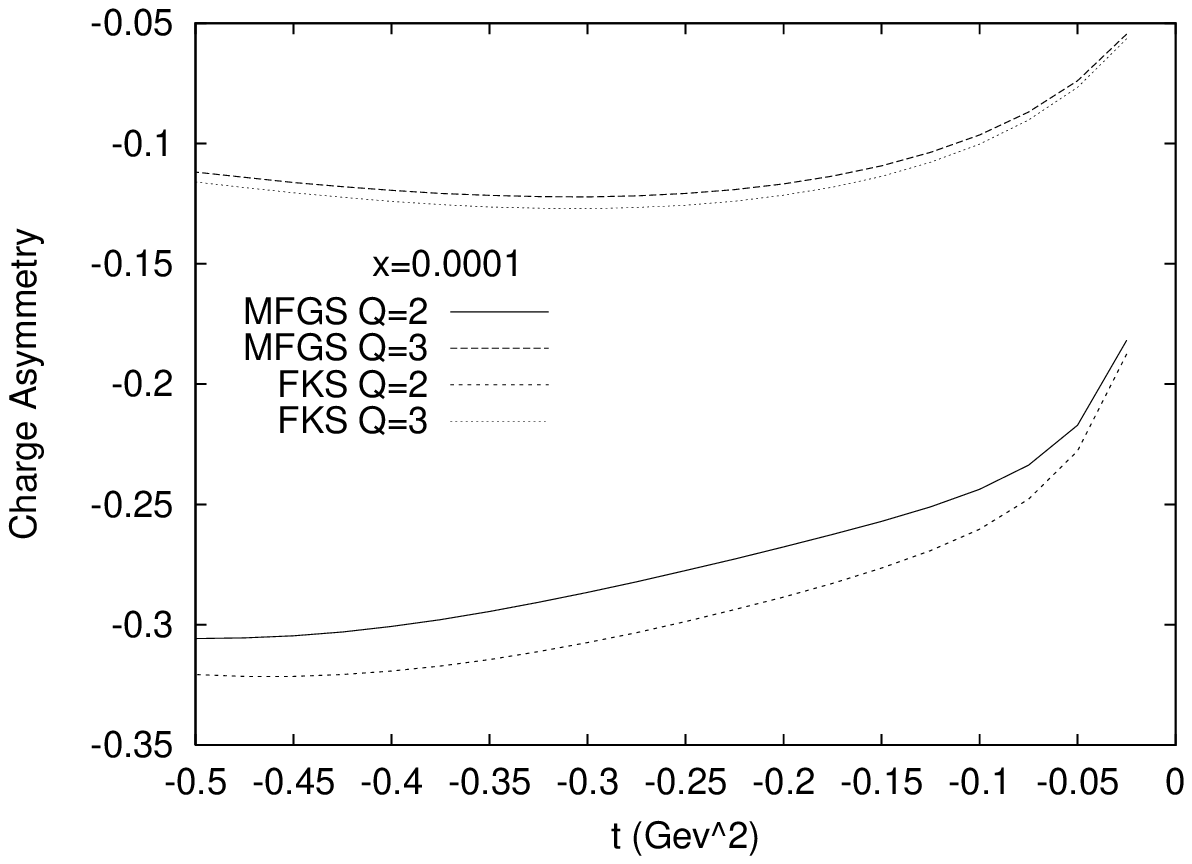}    
\caption{Compendium of results on asymmetries for fixed $x=10^{-4}$, at two values of $Q=2,3$~GeV, accessible in the HERA kinematic range. 
The results for SSA and AAA may be compared to Fig. 1 of \cite{afmm2} and for
CA to Fig. 4a of \cite{bmns2}.}
\label{asymx4}
\end{figure}

\section{Results and conclusions}

\label{sec:res}

The $Q^2$ and $W$ dependencies of the DVCS total cross-section of 
eq.(\ref{sigtot}), obtained from our two dipole models, are compared with 
the H1 data \cite{h1data} in Figs. \ref{fig:photq}, \ref{fig:photw}. 
In these figures, the vertical errors include statistical and systematic errors added in quadrature, and the 
horizontal errors indicate the bin width of the data. 
In making this comparison we have used the same value for the slope parameter, $B = 7$~GeV$^{-2}$,  
as that used by H1 to take into account the resolution and acceptance of their detector. 
It should be emphasized that the uncertainty in $B$ implies an associated uncertainty in the normalization 
of the predictions, which could well be weakly $Q^2$ dependent. 
In addition, Donnachie and Dosch \cite{Heidelberg} have stressed that the 
real photon limit provides an important constraint, since the forward 
imaginary part can be inferred from fits to real photoabsorption data using the optical
theorem, and their resulting value for the real photon cross section $\sigma(\gamma p\rightarrow \gamma p)$
is also shown in Fig. \ref{fig:photw}\footnote{We have actually increased
their estimate by 4 $\%$ to allow for the value of $\beta$ found in the
FKS model. This change is much smaller than the estimated error.}.
As can be seen, the consistency between the predictions and experiment is 
very encouraging for the basic features of the dipole models, despite 
the large statistical errors and the uncertainty in the slope parameter.
Conversely, the close agreement between the predictions of the models out to
energies well beyond the HERA region, despite the fact that one implements 
saturation effects approximately and one does not, means that the total 
cross-section is not a good discriminator between them. This is not unexpected,
since DVCS is clearly a ``softer'' process than DIS, due to the presence of a
real photon.

In figs.(\ref{asymx2},\ref{asymx4}) we show results for the predicted  
asymmetries of eqs.(\ref{aaadef}, \ref{ssadef}, \ref{cadef}) resulting from both models to indicate the overall size 
and spread of predictions expected in the small-$x$ region ($x \in [10^{-2},10^{-4}]$) of HERA kinematics. 
The AAA and SSA results are for a positron and we use HERA kinematics with proton energy 
$E_p  = 920$~GeV (i.e. $S= 4 E_e E_p \approx 99,000$~GeV$^2$) to compute the value of $y = (W^2 + Q^2)/S$.

In conclusion, we observe that both models provide a good description of the available DVCS cross section data, 
without further tuning. We give predictions for the cross section at higher energies. A measurement of the asymmetries 
would allow the predicted phase of DVCS amplitude for both models to be tested.

\section{Acknowledgements}  We are happy to thank A.~Freund and M.~Strikman for
helpful discussions and suggestions. We also thank R.~Stamen and V.~Guzey for helpful comments on the text. The work was supported by PPARC grant number PPA/G/0/1998 and a University of Manchester Research Studentship. M.~M. was also supported by PPARC.


\begin{thebibliography}{100}

\bibitem{zeusdata} P.~R.~Saull, for the ZEUS Collab., ``Prompt photon production and observation of deeply virtual Compton scattering'', Proc. EPS 99 (Tampere, Finland, July 1999), hep-ex/0003030. 

\bibitem{h1data} C.~Adloff {\it et al.}, H1 Collab., Phys.~Lett.~{\bf B517} (2001) 47.

\bibitem{dipole1}  N.~N.~Nikolaev and B.~G.~Zakharov, Z. Phys. {\bf  C49}
(1991) 607; Z. Phys. {\bf C53} (1992) 331; A.~H.~Mueller, Nucl. Phys. 
{\bf B415} (1994) 373; A.~H.~Mueller and B.~Patel, Nucl. Phys. {\bf B425} 
(1994) 471.

\bibitem{levin1} E.~Gotsman, E.~Levin and U.~Maor, Phys. Lett. {\bf B425}
(1998) 369; Eur.~Phys.~J. {\bf C10} (1999) 689; and references therein.

\bibitem{GBW1} K.~Golec-Biernat and M.~W\"{u}sthoff, Phys. Rev. {\bf D59} (1999)
014017; {\bf D60} (1999) 114023.

\bibitem{FKS1} J.~R.~Forshaw, G.~Kerley and G.~Shaw, Phys. Rev. {\bf D60} (1999) 074012.

\bibitem{FKS2} J.~R.~Forshaw, G.~Kerley and G.~Shaw, Nucl. Phys. {\bf A675} (2000) 80c.

\bibitem{FKS3} J.~Forshaw, G.~R.~Kerley and G.~Shaw, ``Colour dipole and saturation'', 
Proc. DIS2000, Liverpool, April 2000, (World Scientific, 2001, p108), 
eds. J.~A.~Gracey and T.~Greenshaw, hep-ph/007257.

\bibitem{MFGS1} M.~McDermott {\it et al.}, Eur. Phys. J. {\bf C16} (2000) 641.

\bibitem{MFGS2} L.~Frankfurt, M.~McDermott, M.~Strikman, JHEP {\bf 0103} (2001) 045.

\bibitem{Heidelberg}  A.~Donnachie and H.~G.~Dosch, Phys. Lett. {\bf B502} (2001) 74.

\bibitem{amirim} M.~F.~McDermott, ``The dipole picture of small $x$ physics (a summary of the Amirim meeting)'', DESY 00-126, hep-ph/0008260.

\bibitem{jcaf} J.~C.~Collins and A.~Freund, Phys. Rev. {\bf D59} (1999) 074009. 

\bibitem{G2} D.~Mueller {\it et al.}, Fortschr. Phys. {\bf 42} (1994) 101;\\
A.~V.~Radyushkin, Phys. Lett. {\bf B380} (1996) 417;\\  
X.~Ji, Phys.~Rev.~Lett. {\bf 78} (1977) 610; Phys.~Rev.~{\bf D55} (1997) 7114.

\bibitem{dglap} V.~N.~Gribov and L.~N.~Lipatov, Sov. J. Phys {\bf 15} (1972) 438, 675; Yu.~L.~Dokshitzer, Sov. Phys. JETP {\bf 46} (1977) 641; G.~Altarelli. and G.~Parisi, Nucl Phys {\bf B126} (1977) 298. 

\bibitem{erbl} A.~V.~Efremov and A.~V.~Radyushkin, Theor. Math. Phys. {\bf 42} (1980) 97, Phys Lett. {\bf B94} (1980) 245; S.~J.~Brodsky and G.~P.~Lepage, Phys Lett. {\bf B87} (1979) 359; Phys. Rev. {\bf D22} (1980) 2157. 

\bibitem{G3} X.~Ji {\it et al.}, Phys. Rev. {\bf D56} (1997) 5511;\\
A.~Radyushkin, Phys. Rev. {\bf D56} (1997) 5554;\\
I.~Musatov and A.~Radyushkin, Phys. Rev. {\bf D61} (2000) 074027.

\bibitem{bmns} A.~Belitsky {\it et al.}, Phys. Lett. {\bf B474} (2000) 163.

\bibitem{bkms} A.~Belitsky {\it et al.}, Phys.~Lett.~{\bf B510} (2001) 117.

\bibitem{ht} N.~Kivel, M.~V.~Polyakov and M.~Vanderhaeghen, Phys. Rev. {\bf D63} (2001) 114014. 

\bibitem{FFS} L.~Frankfurt, A.~Freund and M.~Strikman, Phys. Rev. {\bf D58} (1998)
114001; {\it erratum} {\bf D59} (1999) 119901.

\bibitem{DGS1} A.~Donnachie, J.~Gravelis and G.~Shaw, Eur. Phys. J. {\bf C18} (2001) 539.

\bibitem{afmm1} A.~Freund and M.~McDermott, ``Next-to-leading order evolution of generalized parton distributions for HERA and HERMES'', hep-ph/0106115.

\bibitem{afmm2} A.~Freund and M.~McDermott, ``A next-to-leading order analysis of Deeply Virtual Compton Scattering'', hep-ph/0106124. 

\bibitem{afmm3} A.~Freund and M.~McDermott, ``A next-to-leading order QCD analysis of deeply virtual Compton scattering amplitudes'', hep-ph/0106319. 

\bibitem{rad} A.~Radyushkin, Phys. Rev. {\bf D59} (1999) 014030.

\bibitem{bmns2} A.Belitsky {\it et al.}, Nucl. Phys. {\bf B593} (2001) 289.

\bibitem{DGKP} H.~G.~Dosch {\it et al.}, Phys. Rev. {\bf D55} (1997) 2602. 

\bibitem{FGS}  L.~Frankfurt, V.~Guzey and M.~Strikman,  Phys. Rev. {\bf D58} (1998) 094039.

\bibitem{GVD1}  H.~Fraas, B.J.~Read and D.~Schildknecht, Nucl. Phys. 
{\bf B86} (1975) 346.

\bibitem{GVD2} G.~Shaw,  Phys. Rev. {\bf D47} (1993) R3676; 
G.~Shaw, Phys. Lett. {\bf B228} (1989) 125; 
 P.~Ditsas and G.~Shaw, Nucl.Phys. {\bf B113} (1976) 246.

\bibitem{derivation} L.~Frankfurt, A.~Radyushkin and M.~Strikman,  Phys. Rev. {\bf D55} (1997) 98.

\bibitem{FKopfS} L.~Frankfurt, W.~K\"{o}pf  and M.~Strikman, Phys. Rev. {\bf D54} (1996) 3194; 
Phys. Rev. {\bf D57} (1998) 512.

\bibitem{cteq4l} H.~Lai {\it et al.}, CTEQ Collab.  Phys. Rev. {\bf D55} (1997) 1280.

\bibitem{freund1} A.~Freund and V.~Guzey, Phys. Lett. {\bf B462} (1999) 178; ``Numerical methods in the LO evolution of non-diagonal parton distributions: the DGLAP case'', hep-ph/9801388.

\bibitem{thera} M.~Klein, ``THERA - electron proton scattering at $\sqrt{s} = 1 $ TeV'',  Proc. DIS2000, Liverpool, April 2000, (World Scientific, 2001, p718), eds. J.~A.~Gracey and T.~Greenshaw.\\ 
 
\end{thebibliography}
\end{document}